\documentclass[12pt,preprint]{aastex}

\begin{document}
\title{Afterglow upper limits for four short duration, hard spectrum
gamma-ray bursts\footnote{Partly based on observations collected at
ESO, Chile; Large Program No. 165.H-0464}}

\author{K. Hurley}
\affil{University of California, Berkeley, Space Sciences Laboratory,
Berkeley, CA 94720-7450}
\email{khurley@sunspot.ssl.berkeley.edu}

\author{E. Berger}
\affil{California Institute of Technology, Pasadena CA 91125}

\author{A. Castro-Tirado\altaffilmark{2}}
\affil{Instituto de Astrof\'{\i}sica de Andaluc\'{\i}a (IAA-CSIC),
P.O. Box 03004, E-18080 Granada, Spain}

\author{J. M. Castro Cer\'on}
\affil{Real Instituto y Observatorio de la Armada, Secci\'on de
Astronom\'\i a, 11.110 San Fernando-Naval (C\'adiz) Spain}

\author{T. Cline}
\affil{NASA Goddard Space Flight Center, Code 661, Greenbelt, MD 20771}

\author{M. Feroci}
\affil{Istituto di Astrofisica Spaziale - C.N.R., Rome, I-00133 Italy}

\author{D. A. Frail} 
\affil{National Radio Astronomy Observatory, P.O. Box O, Socorro, NM 87801}

\author{F. Frontera\altaffilmark{3}, N. Masetti}
\affil{Istituto Tecnologie e Studio Radiazioni Extraterrestri, CNR, Via
Gobetti 101, 40129 Bologna, Italy}

\author{C. Guidorzi, E. Montanari}
\affil{Dipartimento di Fisica, Universita di Ferrara, Via Paradiso 12,
44100, Ferrara, Italy}

\author{D. H. Hartmann}
\affil{Clemson University, Department of Physics and Astronomy, Clemson, SC
29634-0978}

\author{A. Henden}
\affil{Universities Space Research Association, US Naval Observatory,
Flagstaff Station, Flagstaff, AZ 86002}

\author{S. E. Levine}
\affil{US Naval Observatory, Flagstaff Station, Flagstaff, AZ 86002}

\author{E. Mazets, S. Golenetskii, D. Frederiks}
\affil{Ioffe Physical-Technical Institute, St. Petersburg, 194021, Russia}

\author{G. Morrison\altaffilmark{4}}
\affil{Caltech-IPAC, M/S 100-22, Pasadena, CA 91125}

\author{A. Oksanen, M. Moilanen}
\affil{Nyr\"ol\"a Observatory, Jyv\"askyl\"an Sirius ry, 
Kyllikinkatu 1, FIN-40100, Jyv\"askyl\"a, Finland }

\author{H.-S. Park}
\affil{Lawrence Livermore National Laboratory, Livermore, CA 94550}

\author{P. A. Price\altaffilmark{5}}
\affil{Palomar Observatory, 105-24, California Institute of Technology,
Pasadena, CA 91125}

\author{J. Prochaska}
\affil{Carnegie Observatories,Headquarters, 813 Santa Barbara St., Pasadena, CA 91101-1292}

\author{J. Trombka}
\affil{NASA Goddard Space Flight Center, Code 691, Greenbelt, MD 20771}

\author{G. Williams}
\affil{Steward Observatory, University of Arizona, Tucson, AZ 85721}

\altaffiltext{2}{Laboratorio de Astrof\'{\i}sica Espacial y F\'{\i}sica 
Fundamental (LAEFF-INTA), P.O. Box 50727, E-28080 Madrid, Spain}

\altaffiltext{3}{Dipartimento di Fisica, Universita di Ferrara, Via Paradiso 12,
44100, Ferrara, Italy}

\altaffiltext{4}{Vanguard Research, Inc., Scotts Valley, CA 95066}

\altaffiltext{5}{Research School of Astronomy and Astrophysics, Mount Stromlo
Observatory, Cotter Road, Weston, ACT 2611, Australia}

\begin{abstract}

We present interplanetary network localization, spectral, and time history information
for four short-duration, hard spectrum gamma-ray bursts, GRB000607, 001025B, 001204, and 010119.
All of these events were followed up with sensitive radio and optical observations
(the first and only such bursts to be followed up in the radio to date),
but no detections were made, demonstrating that the short bursts do
not have anomalously intense afterglows.  We discuss the upper limits, and
show that the lack of observable counterparts
is consistent both with the hypothesis that the afterglow behavior of the 
short bursts is like that of the long duration bursts, many of which
similarly have no detectable afterglows, as well as with the hypothesis that
the short bursts have no detectable afterglows at all.  Small number statistics
do not allow a clear choice between these alternatives, but given the present
detection rates of various missions, we show that progress can be expected
in the near future.

\end{abstract}


\section{Introduction}

It has been recognized for two decades that the time histories of cosmic gamma-ray bursts
appear to fall into at least two distinct morphological categories, namely
the short duration ($\approx$ 0.2 s) bursts, comprising about 25\% of the total,
 and the long duration ($\approx$ 20 s)
bursts , comprising
about 75\% (Mazets et al. 1981; Dezalay et al. 1992; Norris et al. 1984; Hurley et al. 1992; Kouveliotou et al. 1993; Norris et al. 2000).  
The energy
spectra of these two classes of bursts are different: the short bursts tend to
have harder spectra, while the long bursts tend to have softer spectra (Kouveliotou
et al. 1993; Dezalay et al. 1996).  There is also some evidence that their
number-intensity distributions differ (Belli, 1997; Tavani 1998).  However, the two classes 
appear to have identical spatial  (Kouveliotou et al.
1993) and $\rm V/V_{max}$ (Schmidt 2001) distributions. 
Radio
and/or optical counterparts have now been identified for 
a total of about 30 bursts, and spectroscopic
redshifts measured for about 15 of them, but all of them belong to the long duration class.  
Thus,
one of the remaining GRB mysteries is the question of whether the origins of the
long and short bursts are substantially different from one another. 

Over the period 1999 December - 2001 February, the 3rd Interplanetary Network
(IPN) contained two distant interplanetary spacecraft, \it Ulysses \rm and NEAR (the
Near Earth Asteroid Rendezvous mission).  With the near-Earth spacecraft
\it BeppoSAX \rm and Wind (among others), the IPN detected and precisely localized over 100 bursts.
(Prior to 1999 December, and after 2001 February, the IPN had only one distant
spacecraft, \it Ulysses \rm, and produced mainly annuli of location.)  Fifty-six error boxes were produced rapidly and accurately enough to merit rapid
circulation via the GRB Coordinates Network (GCN) circulars, and of these
fifty-six, 34 were searched in the radio, optical, and/or X-ray ranges for
counterparts.  Of the 34 events which were followed up, four were short duration, hard spectrum
bursts with small error boxes.  The IPN localizes bursts by triangulation, or arrival-time
analysis, and in general it derives the smallest error boxes for the
short bursts, since the error box size is directly related to the accuracy
with which the time histories from different spacecraft can be cross-correlated.
A more complete description of the method may be found in Hurley et al. (1999a,b).
We report here on these events and the results of the follow-up searches.  This is the first, and to date the only time that radio observations have
been carried out for this type of burst.  In the optical band, the \it Robotic
Optical Transient Search Experiment \rm (ROTSE-I) has conducted rapid follow-up
observations of three short bursts and obtained magnitude lower limits of $\approx$
13--15 for them (Kehoe et al. 2001); however no deep searches for long-lived optical afterglows have been carried out up to now.  

\section{Gamma-Ray Observations}

Table 1 gives the dates and times of the four bursts, their peak
fluxes and fluences from Konus-Wind measurements, the time interval
over which the peak flux was measured, the spacecraft
which observed them and references to the GCN circulars where they were
announced; the delay between the burst and the issuance of the circulars is
also given, and comments indicate any special circumstances surrounding
the events.  The \it BeppoSAX \rm GRBM did not observe GRB000607 due to
an SAA passage, and was Earth-occulted for GRB001025B.  Figure 1 displays their time histories.  
It is clear from these figures that the four
events fall into the short-duration category.  The fact that all four
bursts were detected as strong events by the NEAR X-Ray/Gamma-Ray Spectrometer (XGRS)
further demonstrates that they have hard energy spectra, because this
experiment has a lower energy threshold $\gtrsim$ 150 keV.  The Konus energy spectra,
shown in figure 2, also show this clearly. 
Instrument references for the IPN experiments may be
found in Hurley et al. (1992 - \it Ulysses \rm GRB), Aptekar et al. (1995 - Konus-Wind), 
 Goldsten et al. (1997), McClanahan et al. (1999) and 
Trombka et al. (1999 - NEAR-XGRS) and Feroci et al. (1997), Amati et al. (1997), 
and Frontera et al. (1997 - \it BeppoSAX \rm GRBM).  

In table 2, the preliminary and final error box areas are
given, as well as the final error box coordinates; the first pair of
coordinates gives the error box center, and the following four pairs,
the corners.  In three cases, as noted,
the final error box is not completely contained within the preliminary one,
due to a larger than usual difference between the preliminary and final
\it Ulysses \rm ephemerides.   This had a minor effect on the observations
reported here.  In those cases where just three spacecraft observed the burst,
the ambiguity between the two triangulated localizations was resolved by the
ability of the Konus-Wind experiment to determine the ecliptic latitude of
the burst.

\section{Follow-up observations}

Attempts were made to detect the optical, infrared, and radio counterparts
to these four bursts; however, no X-ray follow-up observations could be conducted.
Although a \it BeppoSAX \rm target of opportunity program was in place, 
in 3 cases the sources did not satisfy the pointing
constraints, while in the fourth, the delay in deriving the error box
was too long, making the detection of a fading source unlikely.
(We note that one other event, GRB991004,
whose duration was $\sim$ 3.2 s, has been followed up in X-ray observations (in't Zand
et al. 2000); however, this burst could belong to either the short or the long class with
roughly equal probabilities.) 
Table 3 summarizes the optical and IR observations of the final
error boxes.
For each burst, this table gives, in chronological order, the observatory,
the instrument, the delay between the burst and the observation, the band,
the limiting magnitude, the Galactic extinction in the band of observation from
Schlegel et al. (1998) for the low latitude events, the reference to the observation of the
\it initial \rm error box, and any appropriate comments.  (E.g., ``65\% covered'' means that 
only 65\% of the final error box was observed.)  The observations of three of
the bursts were compromised by their proximity to the Sun or Galactic plane.
Further details of the \it Nordic Optical Telescope \rm (NOT) observation of GRB010119
may be found in Gorosabel et al. (2001).
Table 4 similarly summarizes the radio observations, all of which were carried out
with the VLA and covered the entire areas of all the final error boxes.

\section{Discussion}

We now consider the question of whether the radio and optical counterpart
searches were rapid and sensitive enough to have detected counterparts to
these bursts.  In the standard fireball model of the long GRBs (e.g. Wijers, Rees, and Meszaros 1997),
gamma-radiation is produced by internal shocks in the expanding fireball,
while the short- and long-wavelength afterglows are generated when relativistically expanding matter undergoes
external shocks on an ISM which surrounds the source.  There is no correlation, either
in theory or in practice, between the duration of a burst and the decay rate of
its afterglow.  Therefore, in the following, we take as our working hypothesis that the afterglows of the
short bursts are like those of the long bursts, and make no attempt to scale them. 

Between 1997 and 2001, a total of 74 optical and/or IR searches have been carried
out for the counterparts to long duration GRBs, as reported in the literature.  Of them, 50 were
unsuccessful and 24 were successful.
In figure 3, we have characterized these observations by two parameters: the delay in
hours between the burst and the observation, and the detection or upper limit R magnitude.
(In some cases, no R band magnitudes were reported; these events are not plotted.)  
In the same figure, we have similarly characterized and
plotted the upper limits for the four short bursts reported in table 3.  In those cases
where extinction is important, the value of the extinction has been subtracted from
the R magnitude upper limit.  For GRB001025B we have converted the I band upper limits
to R using I-R=0.18, a value which is typical of optical afterglows.

In the same period, as reported in the literature, 14 unsuccessful attempts and 18 successful attempts have
been made to detect the radio counterparts of long GRBs.
Most of these observations were carried out by the VLA at frequencies of 4.86 and 8.5 
GHz.  Detections of the radio counterparts
to the long bursts generally occured at 8.5 GHz or higher frequencies, while the searches
for the short burst counterparts have taken place at 1.43 and 4.86 GHz; at these
lower frequencies, the fluxes of the long bursts tend to be weaker due to synchrotron
self-absorption, but it is not known whether this would similarly affect the observations of
the radio counterparts of the short bursts. In figure 4, we have again characterized each observation by two parameters: the delay
in hours between the burst and the observation, and the detection or upper limit
flux in mJy at 8.5 GHz.  (In some cases, no 8.5 GHz observations were reported;
these events are not plotted.)  We have also plotted the upper limits for the fluxes of the
four short bursts reported in table 4, by assuming a spectral index of -1.5 and
converting the observed upper limits to frequencies of 8.5 GHz.  Thus the upper limits
at 1.43 Gz are increased by a factor of 14.5, and those at 4.86 GHz are increased
by a factor of 2.3.   

Figure 3 demonstrates that the searches for optical and IR counterparts of the short
bursts were generally fast enough and sensitive enough to have detected counterparts,
if we assume that their behavior resembles that of the long bursts.  
That is, counterparts have been detected at roughly the same or later times and/or 
roughly at the same or more intense
fluxes in each case.  Figure 4 similarly shows that 3 out of the 4 radio searches
were fast and sensitive enough to have detected counterparts.
From this we can make a rough prediction
of the results expected from these searches by calculating a ``success rate'' for
counterpart searches.  In the optical, this is $\rm24/74 \sim 32\%$, but this number
should be considered an upper limit, since some unsuccessful attempts may have
gone unreported.  In the radio, numerous unsuccessful attempts have definitely not been
reported, and the actual success rate is $\sim$ 40\%.  
Thus, ignoring possible correlations between the two success rates, 
we would have expected to find $\rm \sim 4 \times 0.32 \, or \, 1.3$ optical
counterparts and, taking into account that only 3 out of 4 of the radio
searches were rapid and sensitive enough, $\rm \sim 4 \times 0.75 \times 0.40 \, or \, 1.2 $radio
counterparts to the four short bursts.  These numbers are consistent with those
actually found, namely 0 and 0, with Poisson probabilities $\sim$ 27 and 30\%, respectively.  
The results of this study are therefore
consistent both with the working hypothesis that the counterparts of the short-duration,
hard-spectrum GRBs behave like those of the long-duration, softer spectrum bursts, as
well as with the hypothesis that the short-duration bursts have no observable counterparts
at all.  (For example, because the fluxes decay more rapidly than those of the long
bursts;
this is considered in Panaitescu et al. 2001.)
Clearly though, the statistics of the small numbers involved, as well as
the difficulties encountered in some of the optical observations, do not 
allow us to choose between these conclusions.

\section{Conclusion}

It has been proposed that extremely brief bursts (those with durations 
$<$100 ms)
may be due to primordial black hole evaporations 
(Cline, Matthey and Otwinowski 1999);
the events which we discuss in this paper have longer durations than this, and the
following considerations therefore do not necessarily apply to them, if
they indeed constitute a separate class. 
Virtually all bursts followed up in X-rays
display X-ray afterglows (Costa 1999), but a large fraction of bursts do not display
detectable long-wavelength afterglows.   It is not known why this is the case, but
possible explanations include sources at very high redshifts, obscured sources, and
very tenuous circum-burster mediums.  The ultimate source of energy for the initial explosion may be
``collapsars'' for the long duration bursts, and merging neutron stars for the short
ones (MacFadyen and Woosley 1999; Ruffert and Janka 1999).  Since a neutron star binary system can
receive a large kick velocity  and subsequently travel far from its host galaxy before
merging (Fryer, Woosley, and Hartmann 1999), short bursts might be expected not to display long-wavelength afterglows,
although they would still have X-ray afterglows (Kumar and Panaitescu 2000).
Thus multi-wavelength afterglow observations hold the key to resolving the short
GRB mystery.  Even if the short bursts are devoid of long-wavelength afterglows, the detection
of X-ray afterglows with \it Chandra \rm or XMM will provide localizations which are precise enough
for deep optical searches to test the host galaxy association.   

Based on the present
data, we can say that the short bursts do not display anomalously intense afterglows
(which we would have detected), but we cannot distinguish the behavior of short bursts from that of the long bursts
with no counterparts.  However, the current interplanetary network, consisting
of \it Ulysses \rm, Mars Odyssey, Konus-Wind, and \it BeppoSAX \rm, is at least
as sensitive as the previous one to short bursts, and it will 
continue to operate for the next several years, as will HETE.  Together they
should provide the data needed to make progress.  For example, after radio observations
of about 12 short bursts have been carried out, the absence of counterparts would be
significant at almost 3$\sigma$ equivalent confidence, and would point to the
conclusion that the afterglows of the short bursts in fact behave differently from those
of the long bursts.

\section{Acknowledgments}

Support for the \it Ulysses \rm GRB experiment is provided by JPL Contract 958056.  
NEAR
data analysis was supported under NASA Grants NAG 5-3500 and NAG 5-9503.  Thanks
also go to Scott Barthelmy for developing and maintaining the GCN, without which most
counterpart searches could not be made.  This research has made use of the NASA/IPAC Extragalactic Database (NED) which is operated by the Jet Propulsion Laboratory, California Institute of Technology, under contract with the National Aeronautics and Space Administration. 
We also thank the staff astronomers at ESO for the observations of GRB000607 and 001204.
The Konus-Wind experiment was supported by RFBR grant \# 99-02-017031 and CRDF grant
\# RP1-2260.

\clearpage

\figcaption{The time histories of the four short bursts from
Konus-Wind.  The Earth-crossing times in seconds of day (UT) corresponding
to time zero on the plots are
8690.4 s for GRB000607, 71366.9 for GRB001025B, 28870.3 s for GRB001204,
and 37178.4 s for GRB010119.
The dashed lines indicate the background levels. \label{Fig. 1}}

\figcaption{The energy spectra of the four short bursts, from
Konus-Wind.  As this instrument has no spectral pre-memory, the
spectra start at the trigger time. \label{Fig. 2}}

\figcaption{Detections of GRB optical counterparts and upper limits.  Each shaded
dot represents an unsuccessful attempt to detect a GRB counterpart, plotted 
according to the
upper limit to its R magnitude and the delay in hours between the burst and the observation.
Each circle similarly represents a successful attempt.  These points have been
taken from the literature, and include IPN, BeppoSAX, and other bursts.  
The numbers 1, 2, 3, and
4 give the magnitude upper limits for the four short bursts (GRB000607, 001025B, 001204, and
010119 respectively) reported in table 3.  For clarity, only the most constraining
points are plotted. \label{Fig. 3}}

\figcaption{Detections of GRB radio counterparts and upper limits.  Each shaded
dot represents an unsuccessful attempt to detect a GRB counterpart, plotted 
according to
the upper limit to its 8.5 GHz flux and the delay in hours between the burst and the observation.
Each circle similarly represents a successful attempt.  These points have been
taken from the literature, and include IPN, BeppoSAX, and other bursts.  The numbers 1, 2, 3, and
4 give the 8.5 GHz flux upper limits for the four short bursts (GRB000607, 001025B, 001204, and
010119 respectively) reported in table 4. For clarity, only the most constraining
points are plotted.\label{Fig. 4}}

\begin{deluxetable}{ccccccccc}
\rotate
\tabletypesize{\scriptsize}
\tablewidth{0pt}
\tablecaption{\it Four short duration, hard spectrum GRBs}
\tablehead{
\colhead{Date}&\colhead{UT at}&\colhead{Instruments or }&\colhead{15-5000 kev}&
\colhead{15-5000 keV} & \colhead{Time interval} & \colhead{Reference}&
\colhead{Delay (h)}&\colhead{Comments} \\
\colhead{(YYMMDD)}&\colhead{Earth, s}&\colhead{spacecraft}&\colhead{fluence, erg cm$^{-2}$} & \colhead{peak flux, erg cm$^{-2}$ s$^{-1}$} & \colhead{ms} & \colhead{}&
\colhead{}&\colhead{}
}
\startdata

000607 &  08690 & \it Ulysses \rm, Konus, NEAR & 5.3 $\rm \times 10^{-6}$ & 1.2 $ \rm \times 10^{-4}$ & 8 & Hurley et al. 2000a & 19.1 & $\rm 35^o$ from Sun \\

001025B & 71346  & \it Ulysses \rm, Konus, NEAR & 5.7 $\rm \times 10^{-6}$ & 2.2 $\rm \times 10^{-5}$ & 16 & Hurley et al. 2000b & 28.9 & $\rm b^{II} \approx 4 ^o$ \\
 
001204 & 28855  & \it Ulysses \rm, Konus, \it BeppoSAX \rm, NEAR & 2.0 $\rm \times 10^{-6}$ & 1.1 $\rm \times 10^{-5}$ & 48 & Hurley et al. 2000c,d &  65.3 & \\

010119 & 37178  & \it Ulysses \rm, Konus, \it BeppoSAX \rm, NEAR & 2.4 $\rm \times 10^{-6}$ & 3.5 $\rm \times 10^{-5}$ & 8 & Hurley et al. 2000e & 14.7 & $\rm b^{II} \approx 5 ^o$ \\

\enddata
\end{deluxetable}

\clearpage
\begin{deluxetable}{cccccc}
\rotate
\tablecaption{\it Error boxes}
\tablewidth{0pt}
\tablehead{
\colhead{Date} & \colhead{Initial area,} & \colhead{Final area,} & \colhead{$\alpha_{2000}$} & \colhead{$\delta_{2000}$} & \colhead{Comments} \\
\colhead{} & \colhead{square arcminutes} & \colhead{square arcminutes} & \colhead{}  & \colhead{}  &
\colhead{}   \\
}
\startdata

000607 & 30 & 5.6 &   2 $^{\rm h}$ 33 $^{\rm m}$ 59.30 $^{\rm s}$  &   17 $^{\rm o}$  8 \arcmin  30.94 \arcsec  & Final error box \\
       &    &     &   2 $^{\rm h}$ 34 $^{\rm m}$  6.88 $^{\rm s}$  &   17 $^{\rm o}$ 10 \arcmin  56.20 \arcsec  & fully contained \\
       &    &     &   2 $^{\rm h}$ 33 $^{\rm m}$ 47.28 $^{\rm s}$  &   17 $^{\rm o}$  3 \arcmin   8.00 \arcsec  & within initial one \\
       &    &     &   2 $^{\rm h}$ 34 $^{\rm m}$ 11.34 $^{\rm s}$  &   17 $^{\rm o}$ 13 \arcmin  54.01 \arcsec  & \\
       &    &     &   2 $^{\rm h}$ 33 $^{\rm m}$ 51.73 $^{\rm s}$  &   17 $^{\rm o}$  6 \arcmin   5.71 \arcsec  & \\

001025B & 110 & 24.5 &  18 $^{\rm h}$ 21 $^{\rm m}$ 23.71 $^{\rm s}$ & -5 $^{\rm o}$  6 \arcmin  23.91 \arcsec  & $\sim$ 3 square arcminutes \\
       &     &      &  18 $^{\rm h}$ 21 $^{\rm m}$ 4.96 $^{\rm s}$  & -5 $^{\rm o}$  4 \arcmin  20.34 \arcsec & outside old \\
       &     &      &  18 $^{\rm h}$ 22 $^{\rm m}$ 3.34 $^{\rm s}$  &  -5 $^{\rm o}$ 13 \arcmin  22.26 \arcsec & error box \\
       &      &     &  18 $^{\rm h}$ 20 $^{\rm m}$ 44.41 $^{\rm s}$ &  -4 $^{\rm o}$ 59 \arcmin  28.15 \arcsec &  \\
       &      &     &  18 $^{\rm h}$ 21 $^{\rm m}$ 42.52 $^{\rm s}$ &  -5 $^{\rm o}$  8 \arcmin  27.90 \arcsec &  \\

001204 & 18  & 6    &  2$^{\rm h}$ 41 $^{\rm m}$ 11.94 $^{\rm s}$ &  12 $^{\rm o}$ 52 \arcmin 
54.3 \arcsec  & $\sim$ 1.3 square arcminutes \\
       &     &      &  2 $^{\rm h}$ 41 $^{\rm m}$ 16.77 $^{\rm s}$ & 12 $^{\rm o}$ 52 \arcmin  14.42 \arcsec  & outside old \\
      &      &      &  2 $^{\rm h}$ 41 $^{\rm m}$  0.39 $^{\rm s}$ & 12 $^{\rm o}$ 51 \arcmin  56.06 \arcsec  & error box \\
      &      &      &  2 $^{\rm h}$ 41 $^{\rm m}$ 23.49 $^{\rm s}$ & 12 $^{\rm o}$ 53 \arcmin  52.58 \arcsec  &  \\   
      &      &      &  2 $^{\rm h}$ 41 $^{\rm m}$ 7.11  $^{\rm s}$ & 12 $^{\rm o}$ 53 \arcmin  34.19 \arcsec  &  \\

010119 & 11 & 3.3 & 18$^{\rm h}$ 53 $^{\rm m}$ 46.17 $^{\rm s}$ &  11 $^{\rm o}$ 59 \arcmin 
47.04 \arcsec  & $\sim$ 1.5 square arcminutes \\
       &     &      & 18 $^{\rm h}$ 53 $^{\rm m}$ 36.00 $^{\rm s}$ & 11 $^{\rm o}$ 59 \arcmin  31.43 \arcsec  & outside old \\
      &      &      & 18 $^{\rm h}$ 53 $^{\rm m}$ 53.61 $^{\rm s}$ & 12 $^{\rm o}$ 00 \arcmin  34.50 \arcsec  & error box \\
      &      &      & 18 $^{\rm h}$ 53 $^{\rm m}$ 39.81 $^{\rm s}$ & 11 $^{\rm o}$ 58 \arcmin  59.57 \arcsec  &  \\   
      &      &      & 18 $^{\rm h}$ 53 $^{\rm m}$ 57.42 $^{\rm s}$ & 12 $^{\rm o}$ 00 \arcmin  02.63 \arcsec  &  \\

\enddata
\end{deluxetable}

\begin{deluxetable}{ccccccccc}
\rotate
\tabletypesize{\scriptsize}
\tablewidth{0pt}
\tablecaption{\it Optical observations}
\tablehead{
\colhead{Date}&\colhead{Observatory}&\colhead{Instrument}& \multicolumn{1}{c}{Delay,}&
\colhead{Band}&\multicolumn{1}{c}{Limiting} & \colhead{Extinction} & \colhead{Reference} & \colhead {Comments} \\
\colhead{    }&\colhead{  }&\colhead{    }&\colhead{h}&
\colhead{    }&\colhead{Magnitude} & \colhead{ } & \colhead{  } & \colhead{ } \\
}
\startdata
000607 & BOOTES-1 & 0.3 m & 51 & R & 16 &  & Masetti et al. (2000) & Near dawn; poor seeing  \\
       & ESO      & 1.54 m & 56 & R         & 19.5\tablenotemark{a}  & & Masetti et al. (2000) & Near dawn; 65\%
coverage \\

001025B & Super-LOTIS & 0.6 m & 30 & unfiltered & 19.5 & 5 & Park et al. (2000) &  \\
       & Calar Alto & 2.2 m & 48 & I & 20.5 & 2.9 & Castro-Tirado et al. (2000) &  \\
       & Las Campanas & 40 in & 52 & I & 21.5 & 2.9  & ... & \\
       & Calar Alto & 2.2 m & 71 & I & 20.5 & 2.9 & Castro-Tirado et al. (2000) & \\
       & Calar Alto & 2.2 m & 96 & I & 20.5 & 2.9 & Castro-Tirado et al. (2000) & \\
       & Super-LOTIS & 0.6 m & 102 & unfiltered & 20.5 & 5 & Park et al. (2000) & \\
       & Las Campanas & 40 in & 109 & I & 21.5 & 2.9  & ... & \\

001204 & USNO        & 1.0 m  & 68  & R$_c$ & 18 & & ... & Heavy clouds, poor seeing \\
       & Mt. Stromlo & 50 in. & 74  & R & 20.1 & & Price et al. (2000) & \\
       & Mt. Stromlo & 50 in. & 74  & V & 20.5 & & ... &  \\
       & ESO         & NTT    & 115 & K$_s$ & 20.0 & & Vreeswijk and Rol (2000) & 80\% coverage \\
       & ESO         & NTT    & 233 & K$_s$ & 20.0 & & Vreeswijk and Rol (2000) & 80\% coverage \\

010119 & NOT         & 2.6 m  & 20 & R & 22.3                & 1.6 & Gorosabel et al. (2001) & \\
       & Palomar     & 60 in. & 27 & R & 18\tablenotemark{a} & 1.6 & Price and Bloom (2000) & Poor seeing \\
       & Nyr\"ol\"a  & 0.4 m  & 42 & R & 19.5 & 1.6 & Oksanen et al. (2001) & Poor seeing; \\
       & Palomar     & 60 in. & 51 & R & 21.5\tablenotemark{a} & 1.6  & Price and Bloom (2000) & 3 \arcsec seeing \\

\enddata
\tablenotetext{a}{Photometric calibration was performed using USNO-A2.0 stars}
\end{deluxetable}

\clearpage

\begin{deluxetable}{ccccc}
\tablewidth{0pt}
\tablecaption{\it NRAO-VLA observations}
\tablehead{
\colhead{Date}&\colhead{Delay, h}&
\colhead{Frequency, GHz}&\colhead{Limiting flux, mJy} & \colhead{Reference}  \\
}
\startdata
000607 &  36 & 1.43 & 0.5  & Frail et al. (2000)   \\
       &  36 & 4.86 & 0.37 & Frail et al. (2000)   \\
       &  40 & 1.43 & 0.5  & Frail et al. (2000)   \\
       &  40 & 4.86 & 0.37 & Frail et al. (2000)   \\
       &  64 & 1.43 & 0.5  & Frail et al. (2000)   \\
       &  64 & 4.86 & 0.37 & Frail et al. (2000)   \\

001025B & 31 & 4.86 & 0.7 & Berger and Frail (2000a)  \\

001204 &  68 & 4.86 & 0.25 & Berger and Frail (2000b)  \\

010119 &  26 & 4.86 & 0.35 & Berger and Frail (2001)   \\

\enddata
\end{deluxetable}

\clearpage

\plotone{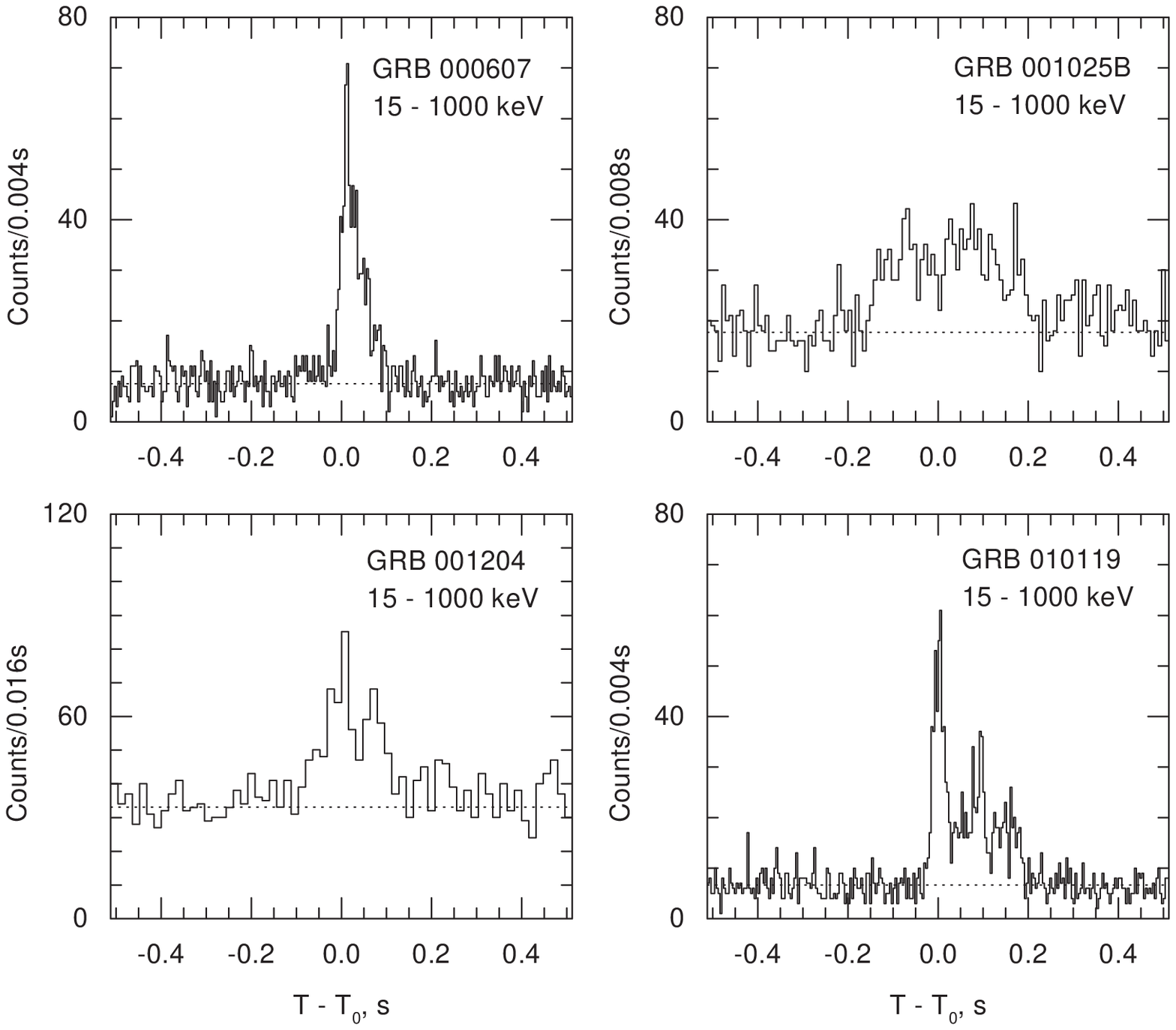}

\clearpage

\plotone{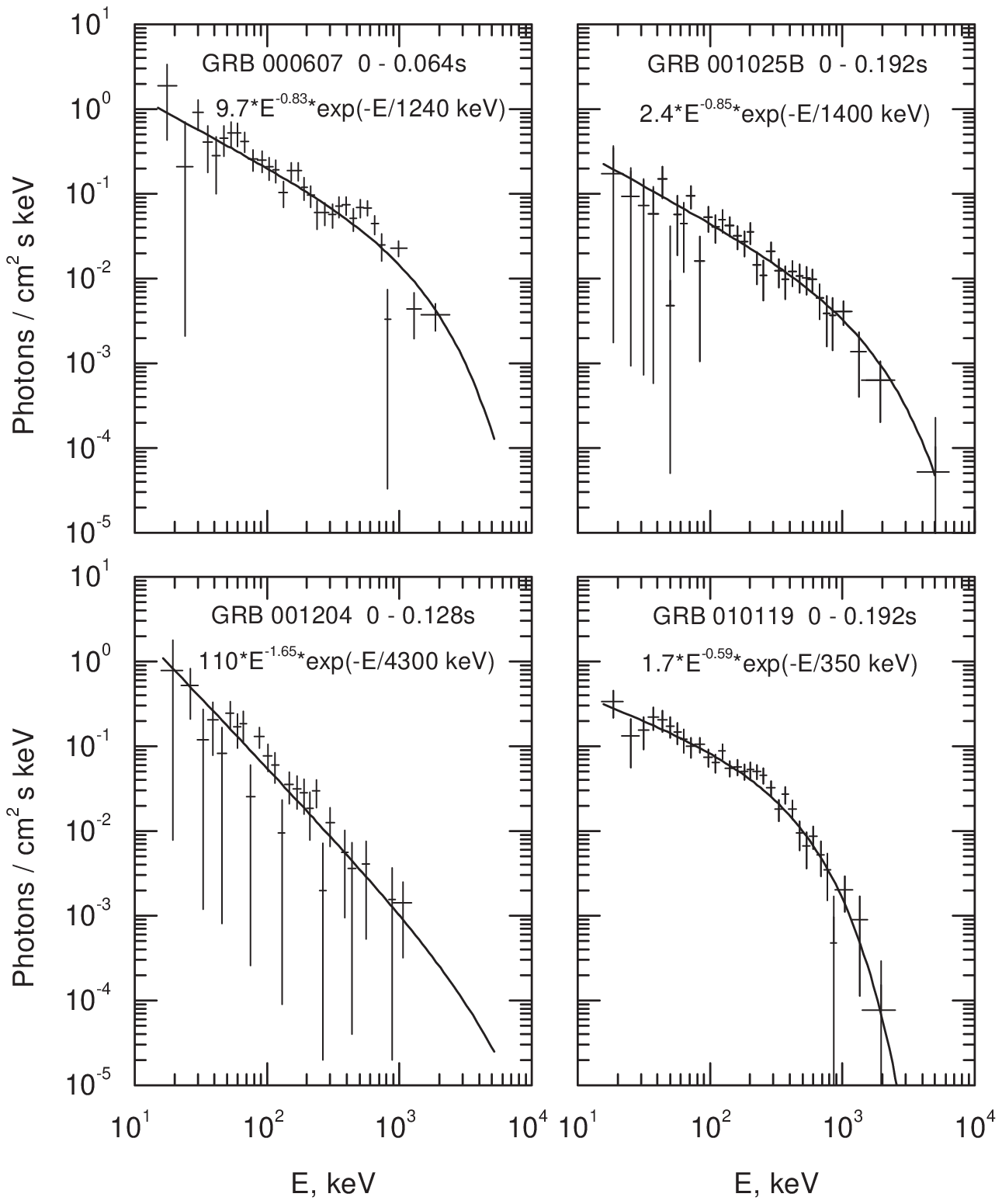}

\clearpage

\plotone{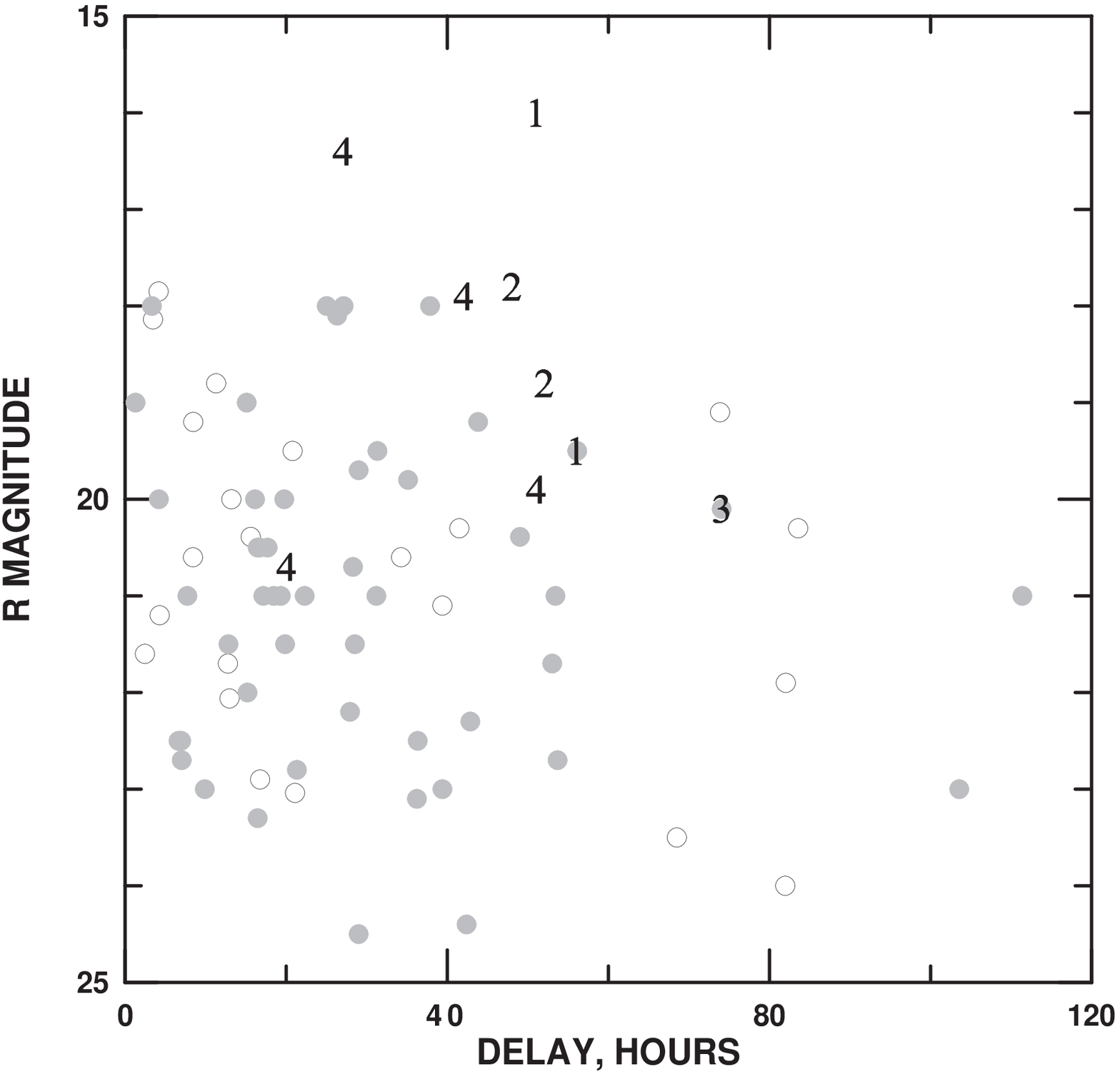}

\clearpage

\plotone{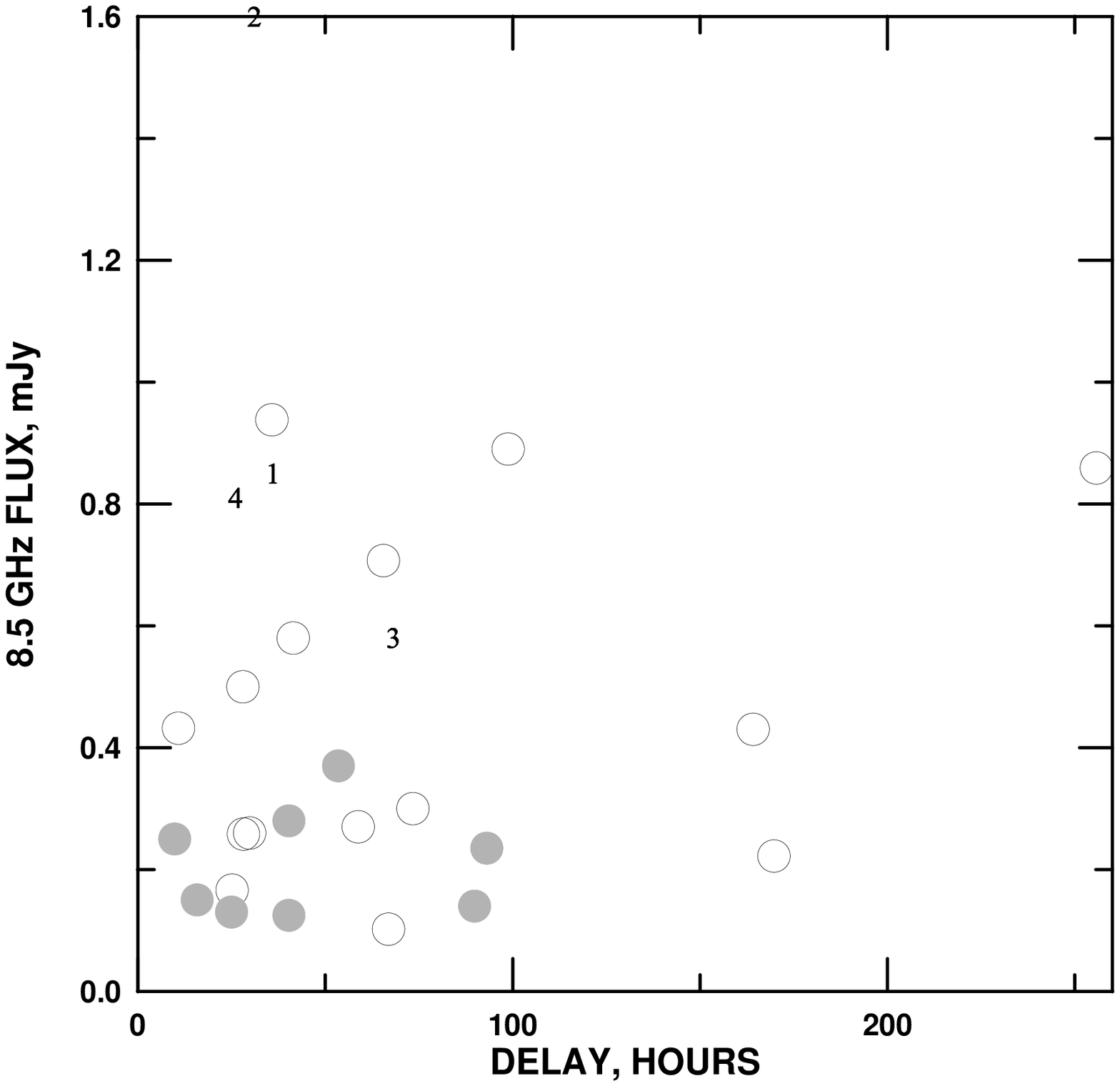}

\clearpage

\end{document}